\documentclass[11pt,french,english,aps,prl,preprint]{revtex4}
\usepackage[T1]{fontenc}
\usepackage[latin9]{inputenc}
\usepackage[a4paper]{geometry}
\usepackage{color}
\usepackage{textcomp}
\usepackage{amstext}
\usepackage{amssymb}
\usepackage{graphicx}
\usepackage{dsfont}

\makeatletter

\providecommand{\tabularnewline}{\\}

\@ifundefined{textcolor}{}
{%
 \definecolor{BLACK}{gray}{0}
 \definecolor{WHITE}{gray}{1}
 \definecolor{RED}{rgb}{1,0,0}
 \definecolor{GREEN}{rgb}{0,1,0}
 \definecolor{BLUE}{rgb}{0,0,1}
 \definecolor{CYAN}{cmyk}{1,0,0,0}
 \definecolor{MAGENTA}{cmyk}{0,1,0,0}
 \definecolor{YELLOW}{cmyk}{0,0,1,0}
}

\usepackage{bbm}
\usepackage{braket}

\makeatother

\usepackage{babel}
\makeatletter
\addto\extrasfrench{%
   \providecommand{\fg}{\ifdim\lastskip>\z@\unskip\fi~\frqq}%
}

\makeatother
\begin{document}
\begin{abstract}
\global\long\def\Ket#1{\ket{#1}}
\global\long\def\Braket#1{\braket{#1}}
\end{abstract}

\title{Stochastic multi-reference perturbation theory with application to
linearized coupled cluster method }

\author{Guillaume Jeanmairet}

\affiliation{Max Planck Institute for Solid State Research, Heisenbergstraße 1,
70569 Stuttgart, Germany}

\author{Sandeep Sharma}

\affiliation{Max Planck Institute for Solid State Research, Heisenbergstraße 1,
70569 Stuttgart, Germany}

\affiliation{Department of Chemistry and Biochemistry, University of Colorado Boulder, Boulder, CO 80302, USA}

\author{Ali Alavi}

\affiliation{Max Planck Institute for Solid State Research, Heisenbergstraße 1,
70569 Stuttgart, Germany}

\affiliation{Department of Chemistry, University of Cambridge, Lensfield Road,
Cambridge CB2 1EW, United Kingdom}
\begin{abstract}
In this article we report a stochastic evaluation of the
recently proposed LCC multireference perturbation theory {[}Sharma
S., and Alavi A., \textit{J. Chem. Phys.} \textbf{143}, 102815, (2015){]}.
In this method both the zeroth order and first order wavefunctions
are sampled stochastically by propagating simultaneously two populations
of signed walkers. The sampling of the zeroth order wavefunction follows
a set of stochastic processes identical to the one used in the FCIQMC
method. To sample the first order wavefunction, the usual FCIQMC algorithm is augmented with a source term that spawns walkers in the sampled first order wavefunction from the zeroth order wavefunction. The second order energy is also computed stochastically but requires no additional overhead outside of the added cost of sampling the first order wavefunction. This fully
stochastic method opens up the possibility of simultaneously treating
large active spaces to account for static correlation and recovering
the dynamical correlation using perturbation theory. 

This method is used to study a few benchmark systems including the carbon dimer and aromatic molecules. We have computed the singlet-triplet
gaps of benzene and m-xylylene. For m-xylylene, which has proved difficult
for standard CASSCF+PT, we find the singlet-triplet gap to be in good agreement with the experimental
values.
\end{abstract}
\maketitle

\section{Intro\label{sec:Intro}}

One of the significant challenges in quantum chemistry is the description
of electronic systems that simultaneously display chemically-relevant
static and dynamical electron correlation. Static correlation often
occurs in open shell systems and gives rise to long-ranged, highly
entangled, many-electron wavefunctions involving electronic orbitals
close to the Fermi energy. Dynamical correlation, on the other hand,
correlates electrons on a short-length scale, and whose descriptions
requires the single-particle spectrum to extend over many energy scales.
There are many systems of practical interest that fall into this class:
we can mention transition metal clusters that are found in many protein
active sites and that play a key role in a wide number of important
biological processes such as photosynthesis or respiration\cite{beinert_iron-sulfur_1997}.
Because of the multideterminental nature of the wave-function they
can prove hard to study with single-reference approaches such as the
widely used density functional theory\cite{neese_critical_2006}.

Unfortunately the computational requirements to describe the wavefunction
of such systems from an exact perspective is utterly daunting. In
an ideal scenario, one would allow for a \textquotedbl{}full\textquotedbl{}
correlation treatment in a large basis: in other words all electrons
(even those not close to the Fermi energy) are simultaneously correlated
over the entire basis. Such a treatment (full CI), which would amount
to the exact solution of the Schrodinger equation in the given basis,
is generally out of reach for sufficient numbers of electrons, owing
the combinatorial explosion of the Hilbert space of a many-particle
system. To overcome this limitation the calculation can be carried
by considering only a meaningful subset of the available configurations,
the most popular choice is to use a complete active space (CAS) wave
function\cite{siegbahn_comparison_1980,roos_complete_1980,siegbahn_complete_1981}. This
approach allows to tackle bigger systems than the FCI one however
it suffers for the same exponential scaling problem. Different
methods have been proposed that allow one to treat larger active spaces by imposing some restrictions on the occupation of the active space orbitals
such as the restricted active space (RAS)\cite{olsen_determinant_1988,malmqvist_restricted_1990},
generalized active space (GAS)\cite{ma_generalized_2011} and SplitGas\cite{li_manni_splitgas_2013}
approaches. With modern techniques such as Density Matrix Renormalization
Group\cite{white_density_1992,white_density-matrix_1993,white_ab_1999,wouters_chemps2:_2014,sharma_spin-adapted_2012,zgid_spin_2008,moritz_convergence_2005,legeza_optimizing_2003,kurashige_high-performance_2009,olivares-amaya_ab-initio_2015}
or the Full CI Quantum Monte Carlo\cite{booth_fermion_2009,cleland_taming_2012,booth_linear-scaling_2013,booth_towards_2013,petruzielo_semistochastic_2012}
technique one can treat very large Hilbert spaces, corresponding to
up to 30-40 electrons in 30-40 orbitals, but even such techniques struggle
to handle systems with many hundreds of electrons correlating in hundreds
or even thousands of orbitals, which is certainly necessary to be able
to treat even intermediate-size molecules. 

Approximations in the correlation treatment are therefore necessary,
and given the energetic divisions between static and dynamical correlation,
multi-reference perturbation theory (MRPT) is a natural starting point, and
which leads to a variety of \textquotedbl{}active-space\textquotedbl{}
methods. Full treatment of correlation among a predefined set of active
orbitals (usually chosen to be around the Fermi energy) with a given
number of electrons leads to the reference Hamiltonian, followed by
a perturbation treatment of the remaining terms in the Hamiltonian
which arise from the remaining terms in the Hamiltonian. In this spirit,
one can mention the complete active space perturbation theory (CASPT)\cite{andersson_second-order_1990,andersson_secondorder_1992},
the n-electron valence state perturbation theory (NEVPT)\cite{angeli_introduction_2001,angeli_n-electron_2002} and the multireference configuration interaction 
(MRCI) method\cite{werner_efficient_1988}, as well as the linearised
coupled cluster (LCC) developed recently by two of us\cite{sharma_multireference_2015,sharma_quasi-degenerate_2016}.
In the perspective of dealing with large systems it is worth noticing
that a second-order perturbation theory approach based on the generalized
active space self-consistent-field has also recently been proposed\cite{ma_second-order_2016}. 

All of these multireference perturbation theories involve a deterministic
resolution of the perturbation equations, and the cost of the calculation
of the perturbation by itself quickly becomes intractable with the
number of core and virtual orbitals. Although this can be dealt by
making a further approximation known as internal contraction, which
comes at the cost of computing the reduced density matrices (RDM)
of the active space up to fourth order, which is itself a significant
challenge for large active spaces. 

A stochastic resolution of the perturbation equation seems to be a
promising direction to reduce the cost of the calculation and to avoid
the use of internal contraction. Surprisingly, only a few attempts
to implement a stochastic resolution of perturbation theories have
been proposed, among which we can mention the stochastic evaluation
of MP2 energies by Hirata and collaborators, MC-MP2. This approach
is based on the rewriting of second order perturbation equations into
the sum of two 13-dimensional integrals thanks to Laplace transform.
Those integrals are then evaluated through Monte-Carlo integration\cite{willow_stochastic_2012,willow_convergence_2013,willow_stochastic_2013}.
Another approach to solve MP2 equations is to express its contributions
in term of graph that describe set of connected Slater determinants,
and then to stochastically sample those graphs\cite{thom_stochastic_2007}.
However those two approaches involve a single reference zero order
wavefunction and to our knowledge stochastic resolution of multireference
perturbation theories have not been proposed yet.

The purpose of this paper is to show how a stochastic treatment of
multireference perturbation theory can implemented within the FCIQMC
methodology, namely a walker-based method to solve for the response
wavefunction of perturbation theory. Response theory differs from
the eigenvalue problem of diagonalisation in that the former involves the solution
of a linear system of equations (the response equations). Since the
FCIQMC technique was developed as a ground-state eigenvalue solver,
it requires to be modified and generalized in order to handle this
new setting. We show how this can be done for the MRLCC method, although
it can be similarly implemented for other flavors of multireference
perturbation theory. Importantly, the resulting method which we term
LCCQMC, is a fully uncontracted method, and is therefore potentially
more accurate than the internally contracted approximations. 

In our new method the sampling of zeroth and first order wavefunctions
are done simultaneously. The population on the zero order wavefunction
follows the standard FCIQMC rules while the population dynamics of
the response wavefunction contains a source term that depends on the
population of the zero order wavefunction. This is practically done
by allowing walkers on one replica to spawn new walkers on another
replica. The structure of the rest of this article is the following,
in the next part we will recall the governing equations of MRPT, with
a particular emphasis on the MRLCC method. We also recall some basics
of the FCIQMC methods before showing how the MRLCC can be expressed
in the \textquotedblleft FCIQMC language\textquotedblright , i.e.
as a stochastic propagation of a signed walkers population. In the
third part a description of the algorithmic of the QMC-LCC method
that has been implemented in the neci code is given, we also discuss
some important technical points. We then illustrate the potential
of the method by applying it to some organic molecules. We first tested
the method by studying the carbon dimer with systematically more refined
basis set, going from cc-pVDZ to cc-pVQZ. Afterwards we turned our
attention to the evaluation of Singlet-Triplet gaps; first in the
case of the benzene molecule which has a singlet for ground state,
and then in the case the m-xylylene diradical that admits a triplet
as its ground state.

\section{Theory\label{sec:Theory}}

\subsection{LCC Perturbation theory}

The essence of quantum-mechanical perturbation theories is to split
the total Hamiltonian $\hat{H}$, into the sum of a simpler Hamiltonian
$\hat{H_{0}}$ and a perturbation operator $\hat{V}$\cite{helhaker_2014},
\begin{equation}
\hat{H}=\hat{H_{0}}+\hat{V},
\end{equation}
with
\begin{equation}
\hat{V}=\hat{H}-\hat{H_{0}}.
\end{equation}
The zero order order energy and wavefunction, $E_{0}$ and $\Ket{\Psi_{0}}$
are solution of the following eigenproblem,
\begin{equation}
\hat{H_{0}}\Ket{\Psi_{0}}=E_{0}\Ket{\Psi_{0}},
\end{equation}
which is assumed to be, and is generally , possible to solve exactly.

In multireference perturbation theories, the zeroth order wavefunction
is expressed as a linear combination of Slater determinants $\Ket{D_{i}}$,
\begin{equation}
\Ket{\Psi_{0}}=\sum_{i}c_{i}\Ket{D_{i}},
\end{equation}
where the expansion set of determinants $\left\{ \Ket{D_{i}}\right\} $
is chosen to recover most of the correlation. Usually the set $\left\{ \Ket{D_{i}}\right\} $
is a CASCI space, it then contains all interactions between active
electrons. 

However the choice of $\hat{H_{0}}$ is not unique since several operators
will admit the CASCI wavefunction as an eigenvector. Among the most
popular ones we can mention the Fock operator used in CASPT\cite{matos_casscfcci_1987,andersson_second-order_1990},
the Dyall\cite{dyall_choice_1995} Hamiltonian used in NEVPT\cite{angeli_introduction_2001,angeli_n-electron_2002}
and the excitation conserving Hamiltonian of Fink\cite{fink_two_2006,fink_multi-reference_2009}
used in the recently proposed MPS-LCC theory\cite{sharma_multireference_2015,sharma_quasi-degenerate_2016}.
Since MRLCC seems to outperformed other methods of similar cost, this
is the one that is used in this study.

If we split the orbitals into an active set where the orbital occupancy
can be 0, 1 or 2, a core set where the orbitals are doubly occupied
and a virtual set of empty orbitals then the total Hamilton, $\hat{H}$
and Fink's Hamiltonian, $\hat{H}_{0}$ can be expressed in second
quantization as,

\begin{equation}
\hat{H}=\sum_{ij}t_{ij}a_{i}^{\dagger}a_{j}+\sum_{ijkl}\Braket{ij|kl}a_{i}^{\dagger}a_{j}^{\dagger}a_{l}a_{k}\label{eq:H@ndQ}
\end{equation}
\begin{equation}
\hat{H_{0}}=\sum_{{ij;\atop \Delta n=(0,0,0)}}t_{ij}a_{i}^{\dagger}a_{j}+\sum_{{ijkl;\atop \Delta n=(0,0,0)}}\Braket{ij|kl}a_{i}^{\dagger}a_{j}^{\dagger}a_{l}a_{k}\label{eq:HFink}
\end{equation}
 where $i,j,k,l$ refer to any orbitals and $\Delta n$ denotes the
change in the total number of electrons between the three subsets
of orbitals. The only operators belonging to $\hat{H_{0}}$ are the
ones that do not transfer electrons between the three subsets. 

The successive correction ($\Ket{\Psi_{m}}$) to the zeroth order
wavefuntion can be computed by using the following equation,
\begin{equation}
\left(\hat{H_{0}}-E_{0}\right)\Ket{\Psi_{m}}=-Q\left(\hat{V}\Ket{\Psi_{m-1}}-\sum_{k=1}^{m-1}E_{k}\Ket{\Psi_{m-k}}\right),\label{eq:bthorderwf}
\end{equation}
 where $Q$ is the projector onto the orthogonal space of the zeroth
order wavefunction. Those sets of equation can be solved sequentially
to compute the $m^{th}$ order of the wavefunction, $\Ket{\Psi_{m}}$.
Once $\Ket{\Psi_{m}}$ is known the $2m$ and $2m+1$ energies can
be computed thanks to Wigner's rules:
\begin{equation}
E_{2m}=\Braket{\Psi_{m-1}|V|\Psi_{m}}-\sum_{k=1}^{m}\sum_{j=1}^{m-1}E_{2m-k-j}\Braket{\Psi_{k}|\Psi_{j}},
\end{equation}
\begin{equation}
E_{2m+1}=\Braket{\Psi_{m}|V|\Psi_{m}}-\sum_{k=1}^{m}\sum_{j=1}^{m}E_{2m+1-k-j}\Braket{\Psi_{k}|\Psi_{j}}.
\end{equation}

Note that with the definition of the zeroth order Hamiltonian given
in Eq.\ref{eq:HFink}, the first order energy $E_{1}=\Braket{\Psi_{0}|V|\Psi_{0}}$
vanishes.

In a recent paper\cite{sharma_multireference_2015}, both the 0 order
wavefunction and the successive corrections were expressed as matrix
product states (MPS) and computed deterministically by functional
minimization. Here we proposed an alternative approach where both
the zero order and the perturbation wavefunctions are sampled stochastically
in the Fock space, the perturbation second-order energy is also evaluated
stochastically.

To sample the zero order wavefunction and energy we used the standard
FCIQMC approach restricted to the CAS space. We will recall here the
main points of this approach. 

\subsection{Elements of FCIQMC}

FCIQMC is a method which aims at stochastically minimizing the energy
of a ground state wavefunction expressed as a CASCI (or Full-CI) expansion.
The wavefunction can be expressed as a linear combination of determinants
belonging to the CAS-CI space.
\begin{equation}
\Ket{\Psi}=\sum_{i}c_{i}\Ket{D_{i}}.\label{eq:expansionWF}
\end{equation}

Formally, the idea is to find the ground state of the Hamiltonian
operator $\hat{H}$, by integrating the imaginary time Schrodinger
equation (ITSE),
\begin{equation}
\frac{\partial\Ket{\Psi}}{\partial\tau}=-\hat{H}\Ket{\Psi}.\label{eq:imtimeSE}
\end{equation}
The discretization of Eq.\ref{eq:imtimeSE} with a time step $\Delta\tau$
leads to the following evolution equation
\begin{equation}
\Ket{\Psi(t+\Delta\tau)}=\left(\mathds{1}-\Delta\tau\left(\hat{H}-S\mathds{1}\right)\right)\Ket{\Psi(t)}.
\end{equation}

$S$ is a shift parameter used to control the walkers population and
$\mathds{1}$ is the identity operator. Thus starting from a guess
wavefunction, for instance the Hartree Fock determinant, the ground
state can be reached by repetitively applying the following projector
. 
\begin{equation}
\hat{P}=\mathds{1}-\Delta\tau\left(\hat{H}-S\mathds{1}\right),\label{eq:projector}
\end{equation}
To circumvent the prohibitive storage of the full CI vector, in FCIQMC
this projection operation is realized scholastically such as the proper
projection is recovered on average. To do so the coefficient of the
expansion in Eq.\ref{eq:expansionWF} are sampled by a population
of signed walkers. Each of those carries a signed weight and is located
on a Slater determinant. The total signed sum of the walkers residing
on the same determinant can be interpreted as an instantaneous measure
of its weight $c_{i}$. The walker population evolves through a set
of stochastic processes that mimic the projector of Eq.\ref{eq:projector}:
\begin{enumerate}
\item A cloning/death step, in which the walker population on each determinant
is increased/reduced with a probability $\left(H_{ii}-S\right)\Delta\tau$.
$S$ is a shift parameter that is used to control the total walker
population.
\item A spawning step. For each walker on a determinant $\Ket{D_{i}}$ a
singly or doubly connected determinant $\Ket{D_{j}}$ is generated
with a probability $p_{gen}^{ij}$. A signed child is actually generated
on the determinant $\Ket{D_{j}}$ with a spawning probability 
\begin{equation}
p_{spawn}^{ij}=\frac{\left|H_{ij}\right|\Delta\tau}{p_{gen}^{ij}}.\label{eq:pspawn}
\end{equation}
The sign of the newly spawned walker is the same as the sign of the
parent if $H_{ij}>0$, it is of opposite sign otherwise.
\item Each pairs of negative and positive newly spawned walkers lying on
the same determinant are removed during an Annihilation step. This
avoid the growth of an infinite noise due to the so-called sign problem.
\end{enumerate}
We propose here a modification of the FCIQMC algorithm in order to
stochastically sample simultaneously the zeroth order wavefunction
and the successive order of the perturbation of Eq.\ref{eq:bthorderwf}.
Even if in principle any order of perturbation can be reached by this
technique, in this article we will only consider the calculation of
the fist order correction to the wave function, that is given through
Eq.\ref{eq:bthorderwf} by 
\begin{equation}
\Ket{\Psi_{1}}=\left(\hat{H}_{0}-E_{0}\right)^{-1}Q\hat{V}\Ket{\Psi_{0}}\label{eq:ps1equ}
\end{equation}

In that case the problem is simpler since the Hilbert space on which
the zeroth and first order wavefunctions are expanded is limited to
the CASSD space. Moreover this space can be expressed as a direct
sum of two subspaces ${\cal H}={\cal H}_{0}\oplus{\cal H}_{1}$, where
${\cal H}_{0}$ correspond to the CAS space and ${\cal H}_{1}$ is
its orthogonal compliment which contains all the determinants that
are single or double excitations from the ones belonging to ${\cal H}_{0}$.
Applying $\hat{V}$ to $\Ket{\Psi_{0}}$ only generates determinants
on ${\cal H}_{1}$. There is thus no need to ensure the orthogonality
of the two wavefunctions and the $Q$ projector operator in Eq.\ref{eq:ps1equ}
can be dropped. Of course higher order wavefunctions would contain
determinants that are higher order excitations from the CASCI space.
Orthogonalization with respect to $\Ket{\Psi_{0}}$ and to the lower
order perturbation wavefunctions would also be required. 

The zero order wavefunction follows an ITDSE similar to Eq. \ref{eq:imtimeSE},
\begin{equation}
\frac{\partial\Ket{\Psi_{0}}}{\partial\tau}=-\hat{H}_{0}\Ket{\Psi_{0}},\label{eq:imtimeSEH0}
\end{equation}
it thus can be solved by using the standard FCIQMC algorithm, with
the exception that new generated determinants will be restricted to
the one accessible by applying $\hat{H_{0}}$ on the current wave-function.
This will effectively restrict the Hilbert space to the CAS space,
and correspond to freezing the core and virtual orbitals. The computation
of $\Ket{\Psi_{1}}$ as defined in Eq.\ref{eq:ps1equ} is less straighforward.
Indeed in FCIQMC we do not have access to a proper description of
the zeroth order wavefunction. It is thus not possible to compute
$\Ket{\Psi_{1}}$ by using a projection approach. As an alternative
we also decide to sample $\Ket{\Psi_{1}}$ stochastically. However
as opposed to $\Ket{\Psi_{0}}$ the perturbation wavefunctions are
not solution of an ITDSE. We introduced the following hierarchy of
differential equation (DE). 

\begin{equation}
\frac{\partial\Ket{\Psi_{m}}}{\partial\tau}=-\left(\hat{H}_{0}-E_{0}\right)\Ket{\Psi_{m}}-QV\left(\Ket{\Psi_{m-1}}-\sum_{k=1}^{m-1}E_{k}\Ket{\Psi_{m-k}}\right).\label{eq:pismDestoch}
\end{equation}
If the left hand side of Eq.\ref{eq:pismDestoch} cancels out, we
recover the expression of Eq.\ref{eq:bthorderwf} for $\Ket{\Psi_{m}}$,
in other words the successive correction wavefunctions are stationary
solution of these DE. Note that propagating this equation is actually
going to reach a steady state, that is equal to $\Ket{\Psi_{m}}$
because the $\left(\hat{H}_{0}-E_{0}\right)$ matrix is positive definite.
In particular the DE for the first order perturbation is
\begin{equation}
\frac{\partial\Ket{\Psi_{1}}}{\partial\tau}=-\left(\hat{H}_{0}-E_{0}\right)\Ket{\Psi_{1}}-V\Ket{\Psi_{0}}.\label{eq:psi1diffeq}
\end{equation}

This equation is similar to the one used for $\Ket{\Psi_{0}}$ , with
the addition of a source term due to the second term of the right
hand side. In the next section we will show how the simultaneous solving
of Eq.\ref{eq:imtimeSEH0} and Eq.\ref{eq:psi1diffeq} has been implemented
in the NECI program.

\section{Implementation\label{sec:Implementation}}

To simultaneously sample the zeroth and first order wavefunctions,
we use the multi-replica technique\cite{overy_unbiased_2014}. A first
replica, labeled 0, is sampling the 0 order wavefunctions by propagating
the ITDSE of Eq.\ref{eq:imtimeSEH0} while another one, labeled 1,
is sampling the first order perturbation. We first start with a small
amount of walkers on a reference determinant, typically the Hartree-Fock
one, on replica 0 and no walkers on replica 1.

In replica 0 new walkers are spawned by applying one and two electrons
operators that belong to $\hat{H}_{0}$, thus only determinants that
belong to the CAS space are generated. Because the sampling of $\Ket{\Psi_{0}}$
is equivalent to a standard FCIQMC sampling in a CASCI, we can use
all the optimizations and approximations that have been introduced
in previous publications such as initiator approximation\cite{cleland_communications:_2010,cleland_taming_2012}
or the semi-stochastic approximation\cite{blunt_semi-stochastic_2015}.

Once the population on replica 0 is equilibrated we start to sample
$\Ket{\Psi_{1}}$. This equilibration of the zeroth order wavefunction
can be monitored by looking at the variational energy for this replica
and checking that is correspond to the CASCI energy. At this point
we attempt multiple spawning from each walkers of replica 0. In addition
to the excitation that belong to $\hat{H}_{0}$ we also generate excitation
belonging to $\hat{V}$. This thus generates determinants that belong
to the external space, ${\cal H}_{1}$. The spawning probability of
those $\Ket{\Psi_{0}}$ to $\Ket{\Psi_{1}}$ walkers follows the expression
of Eq.\ref{eq:pspawn}. However those walkers are spawned on replica
1 instead of replica 0. 

Replica 1 that was initially empty starts getting populated, the walkers
coming from replica 0 correspond to the source term in Eq.\ref{eq:psi1diffeq}.

\begin{figure}
\centering{}\includegraphics[width=0.6\textwidth]{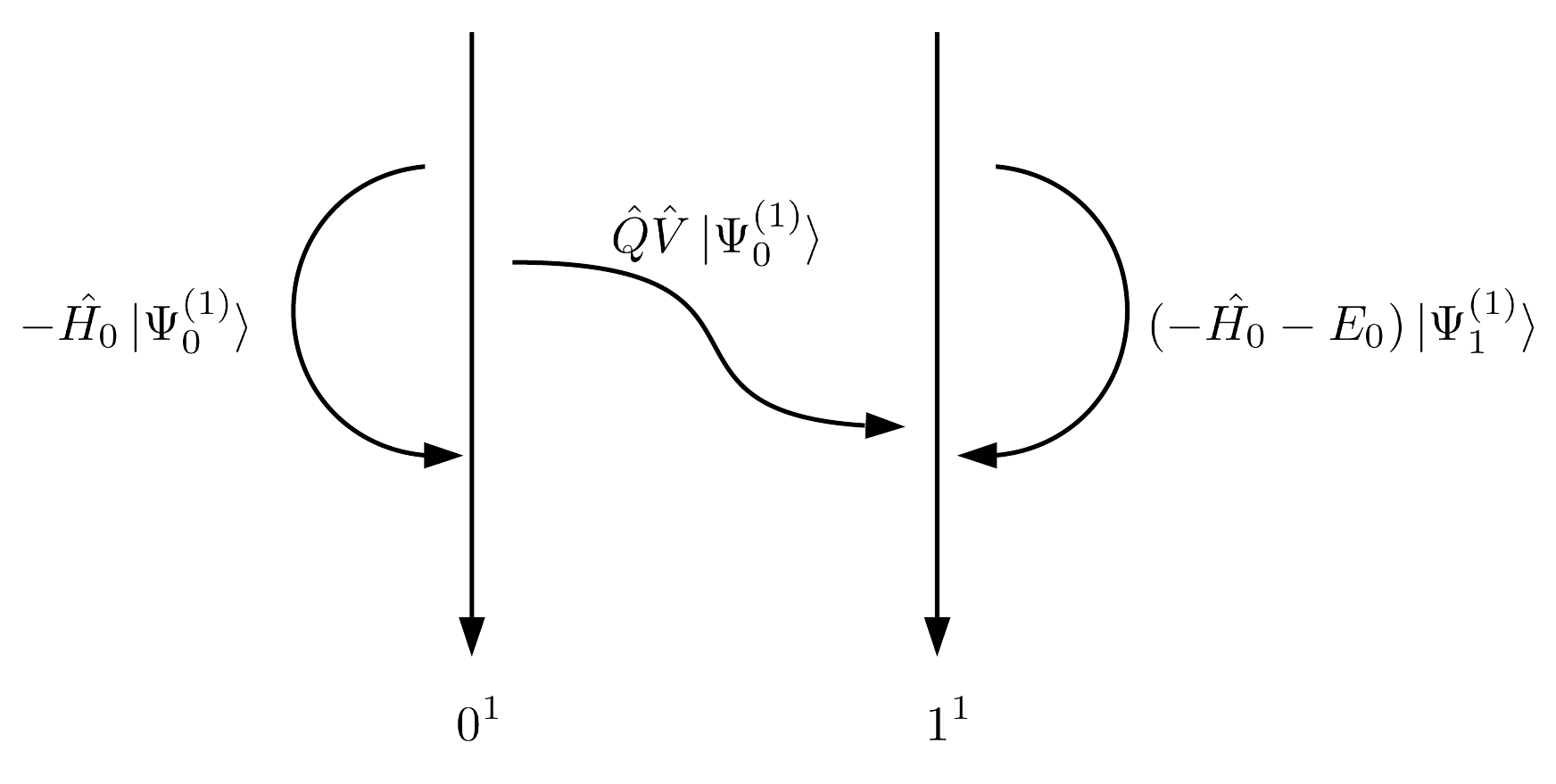}\caption{Schematized description of an iteration update of the two replicas.
On the left the 0 order wave function, sampled in replica 0 is updated
by applying the $\left(\mathds{1}-\Delta\tau\left(\hat{H_{0}}-S\mathds{1}\right)\right)$
operator. On the right the first order perturbation, in replica 1,
is updated by applying the $\left(\mathds{1}-\Delta\tau\left(\hat{H_{0}}-E_{0}\mathds{1}\right)\right)$
operator to $\protect\Ket{\Psi_{1}}$ and adding walkers spawned by
applying $\hat{V}$ onto $\protect\Ket{\Psi_{0}}$. \label{fig1:scheme}}
\end{figure}

We emphasize the fact that in replica 0 the population dynamic is
not modified by this extra excitation step, and that by construction
the first and zeroth order wavefunctions are orthogonal to each other,
this obviates the use of orthogonalization techniques that will be
required for the higher order perturbation. The walkers on replica
1 are also subjected to a cloning/dying step and a spawning step at
each iteration. For the dying step the applied operator is $\left(\hat{H}_{0}-E_{0}\right)$,
where the $E_{0}$ is the projected energy in replica 0. For the spawning
step, the applied excitation operator belongs to $\hat{H}_{0}$. The
simultaneous sampling of the two wavefunctions is schematized in Fig.\ref{fig1:scheme}.

Note that with this implementation the timesteps used for $0\rightarrow0$,
$0\rightarrow1$ and $1\rightarrow1$ spawning steps, and the 0 and
1 cloning and dying steps are identical. It is the one that has been
optimized for replica 0. In practice this timestep is chosen to ensure
that the spawning probability in replica 0 is small enough to ensure
that the spawning probability is not much bigger than 1 for all $\Ket{D_{i}}$.
As can be seen of Eq.\ref{eq:pspawn} the spawning probability is
inversely proportional to the generation probability $p_{gen}$ to
attempt a spawning on $\Ket{D_{j}}$ from $\Ket{D_{i}}$. Because
in interesting systems the size of the external space is expected
to be much bigger than the one of the active space the probability
of generating $\Ket{D_{i}}$, $\Ket{D_{j}}$ pairs is much smaller
in case of $0\rightarrow1$ and $1\rightarrow1$ than in case of $0\rightarrow0$
excitation. As a consequence the spawning probabilities of those excitation
will be much bigger than $0\rightarrow0$ spawning probability for
the same timestep, and a single walker would give birth to multiple
walkers; such an event is called blooming. In other words, the time
step should be smaller in the excitations involving the response functions.
This problem is dealt with as follow, the overall time step of the
simulation is set by the $0\rightarrow0$ dynamic. The first time
blooming occurs during the $0\rightarrow1$ spawning step we keep
track of the biggest bloom and compute the first integer bigger than
this, $n_{01}$. During further step, the time step value will be
divided by this $n_{01}$ factor for $0\rightarrow1$ spawning, this
ensures to have a spawning probability not much bigger than 1. To
keep the overall dynamic at the same timestep we have to do $n_{01}$
spawning attempt to $\Ket 1$ for each walkers on $\Ket 0$. We use
a similar procedure for the $1\rightarrow1$ spawning defining a $n_{11}$
timestep scaling factor. Those $n_{01}$ and $n_{11}$ factors are
updated along the run, this prevent any explosion of the replica 1
population.

\begin{figure}
\centering{}\includegraphics[width=0.7\textwidth]{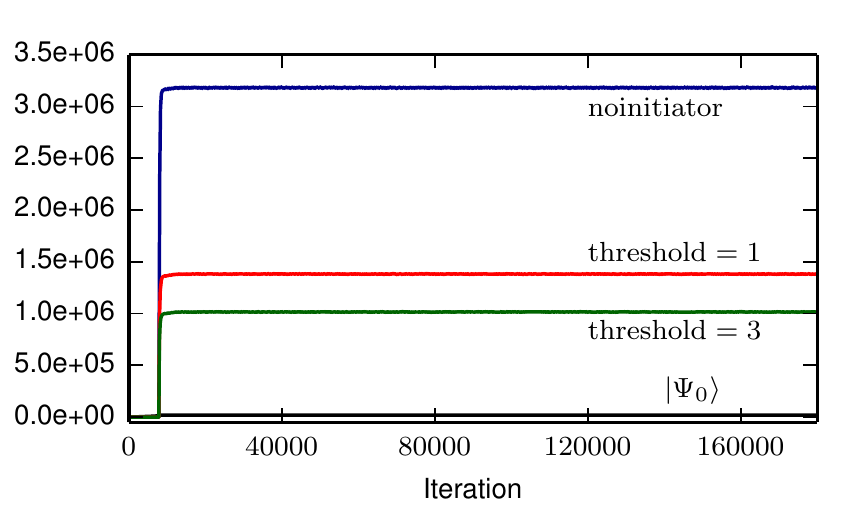}\caption{Number of walkers in replica 0 (black), and in replica 1 with an initiator
approximation of 3 (green), 1 (red) and no initiator approximation
(blue) for the $C_{2}$ molecule in the cc-pVQZ basis set. \label{fig1:swalknum}}
\end{figure}

With this implementation there is no way to control the total number
of walkers on $\Ket{\Psi_{1}}$ because there is no analogue of the
shift control parameter that is used for replica 0. After a few steps
the total number of walkers in replica 1 reaches a plateau that is
dependent of the system. As a first attempt to control the value of
that plateau the initiator approximation has been implemented for
walkers that belong to replica 1. We allow the initiator threshold
to be different in replica 0 and replica 1. As an illustration we
studied the carbon dimer molecule with the cc-pVQZ basis set\cite{jr_gaussian_1989},
with ${\cal H}_{0}$ corresponding to a CAS (8,8) which is the valence
space of the molecule. We use a initiator threshold of 3 and a targeted
number of walker of 50k for replica 0. In Fig.\ref{fig1:swalknum}
we present the evolution of the number of walker on replica 0 and
on replica 1, the different calculations have been run with a initiator
criterion of 3, 1 and no initiator approximation on replica 1.

Looking at Fig.\ref{fig1:swalknum} it can be seen that with no initiator
approximation on $\Ket{\Psi_{1}}$, the number of walkers grows to
more than 60 times the number of walkers in the reference. When using
the most moderate initiator criteria of 1, this number is already
reduced by more than a factor of 2, it can be further decreased by
increasing the initiator threshold. However going from a threshold
of 1 to 3 only reduced the total number of walkers by roughly 30 \%.
This is still not satisfying since there is no way to know \textit{a
priori} what the number of walkers on $\Ket{\Psi_{1}}$ is going to
be. The cost of the calculation cannot be known before running it;
for this reason we describe hereafter a way to control the population
on replica 1 independently from the initiator threshold.

In Eq.\ref{eq:ps1equ} it can be seen that $\Ket{\Psi_{1}}$ scales
linearly with the perturbation $\hat{V}$. \textcolor{black}{We start
by making the assumption that the number of walkers on $\Ket{\Psi_{1}}$
also depends linearly on the perturbation} and thus scale down the
perturbation by a real prefactor $\alpha$, that is typically small.
This is done in practice by multiplying the matrix elements $H_{ij}$
by this factor when the spawning probability from a determinant $\Ket{D_{i}}$
on ${\cal H}_{0}$ to a determinant $\Ket{D_{i}}$ on ${\cal H}_{1}$
is computed. This allows us to tune more easily the total number of
walkers on replica, also the relation between the value of the plateau
and the $\alpha$ remains unknown. The number of walkers does not
strictly scales linearly with the value of $\alpha$ \textcolor{black}{because
of the offdiagonal spawning in replica 1} and because of the initiator
approximation. To circumvent this problem we implemented a dynamic
updating of $\alpha$ in order to reach a target number of walker
on $\Ket{\Psi_{1}}$. The simulation is started with a small $\alpha$,
typically $10^{-2}$ ; after the a few thousand step of equilibration,
if the number of walkers on replica 1 is not included within a 3\%
treshold of the target, we update alpha to a new value $\alpha^{\prime}$,
\begin{equation}
\alpha^{\prime}=\alpha\left(\gamma+(1-\gamma)\frac{N_{t}}{N}\right)
\end{equation}
, where $N_{t}$ is the target number of walkers, $N$ is the current
number of walkers, and $\gamma$ is a dumping parameter to prevent
too drastic a change of $\alpha$. We use typically $\gamma=0.5$. 

Having implemented this way of controlling the population in replica
1 we now will study the influence of the number of walkers used to
sample the response function on the second order energy. The second
order correction energy is expressed as 
\begin{equation}
E_{2}=\frac{\Braket{\Psi_{0}|\hat{V}|\Psi_{1}}}{\Braket{\Psi_{0}|\Psi_{0}}}=\frac{\sum_{i\in{\cal H}_{0}}\sum_{j\in{\cal H}_{1}}c_{i}c_{j}\Braket{D_{i}|\hat{V}|D_{j}}}{\sum_{i\in{\cal H}_{0}}ci^{2}}.
\end{equation}
In the FCIQMC framework this can be rewritten as a function of the
walker population on each determinants involved. In the LCC pertrubation
theory $\Braket{D_{i}|\hat{V}|D_{j}}=\Braket{D_{i}|\hat{H}|D_{j}}=H_{ij}$
\begin{equation}
E_{2}\approx\left\langle \tilde{E}_{2}\right\rangle =\frac{\Braket{\Psi_{0}|\hat{V}|\Psi_{1}}}{\Braket{\Psi_{0}|\Psi_{0}}}=\frac{\left\langle \sum_{i\in{\cal H}_{0}}\sum_{j\in{\cal H}_{1}}N_{i}N_{j}H_{ij}\right\rangle }{\left\langle \sum_{i\in{\cal H}_{0}}N_{i}^{2}\right\rangle }=\frac{\left\langle \sum_{j\in{\cal H}_{1}}N_{j}\left[\sum_{i\in{\cal H}_{0}}N_{i}H_{ij}\right]\right\rangle }{\left\langle \sum_{i\in{\cal H}_{0}}N_{i}^{2}\right\rangle },\label{eq:E2}
\end{equation}
\textcolor{black}{where $\left\langle .\right\rangle $ denotes the
average over multiple timesteps and $\tilde{E}_{2}$ is second order
energy computed with the instantaneous population in the two replica. }

\textcolor{black}{In principle we could exactly compute $\tilde{E}_{2}$
by finding all the connections between determinants belonging to ${\cal H}_{0}$
and ${\cal H}_{1}$ but this is prohibitive. Instead the connection
are found} through the spawning process, this strategy has already
been used for the calculation of reduced density matrices\cite{overy_unbiased_2014}.
When a successful spawning from a determinant $\Ket{D_{i}}$ in replica
0 to a determinant $\Ket{D_{j}}$ in replica 1 occurs, the product
of the matrix element and of the number of walkers on $\Ket{D_{i}}$,
$N_{i}H_{ij}$ is communicated along the spawned walker to the processor
holding the child determinant $\Ket{D_{j}}$. Once all the spawning
attempt have been done, the processor that keep track of the $\Ket{Dj}$
walkers population will also contains all the $N_{i}H_{ij}$ contribution
from all the determinants that spawned to $\Ket{Dj}$ this iteration.
This strategy does not cause any noticeable increase of the computational
cost. As the contribution of a $\Ket{D_{i}},\Ket{D_{j}}$ pair of
determinants to the $E_{2}$ energy is only taken into account when
a successful spawning step is actually happening this contribution
should be rescaled by the normalized probability of spawning at least
one child (of any weight) onto $\Ket{D_{j}}$ from $\Ket{D_{i}}$
during the current iteration. More details on how to compute this
probability can be found in \cite{overy_unbiased_2014}. To avoid
double counting of $C_{i}C_{j}$ contribution, it is necessary to
carefully check for the rare but still possible case of multiple spawning
from the same determinant $\Ket{D_{i}}$ to the same determinant $\Ket{D_{j}}$.
Thus the $N_{i}H_{ij}$ contribution is only communicated to the processor
holding $\Ket{D_{j}}$ in the first occurrence of such an event.\textcolor{red}{{} }

\textcolor{black}{An additional difficulty comes from the computation
of the denominator of Eq.\ref{eq:E2}. If we just take the square
of the number of walkers on a determinant in replica 0, a bias is
introduced because $\left\langle N_{i}\right\rangle \left\langle N_{i}\right\rangle \neq\left\langle N_{i}^{2}\right\rangle $,
the square of the instantaneous value of the walker population is
correlated. In the RDM computation this particular problem of the
normalization was circumvent by using the trace condition to normalize
RDM }\textit{\textcolor{black}{a posteriori}}\textcolor{black}{. There
is no such relation to unbias the measure of the norm of $\Ket{\Psi_{0}}$
here, instead we use a replica trick. This requires the addition a
third replica, labeled $0^{\prime}$ that is also sampling the zero
order wavefunction. Because the two replica $0$ and $0^{\prime}$
are uncorrelated we can have an unbiased measure of the norm of $\Ket{\Psi_{0}}$
by $\Braket{\Psi_{0}^{(0)}|\Psi_{0}^{(0^{\prime})}}=\sum_{i\in{\cal H}_{0}}N_{i}^{(0)}N_{i}^{(0^{\prime})}$.}
Finally, it is necessary to rescale the $\Ket{\Psi_{1}}$ function
by the $\alpha$ factor to compute the $E_{2}$ energy. 

With this efficient way of computing the second order energy, we can
go back to the example of Fig. \ref{fig1:swalknum} and examine how
the estimation of $E_{2}$ converge with the initiator approximation.

\begin{figure}
\centering{}\includegraphics[width=0.8\textwidth]{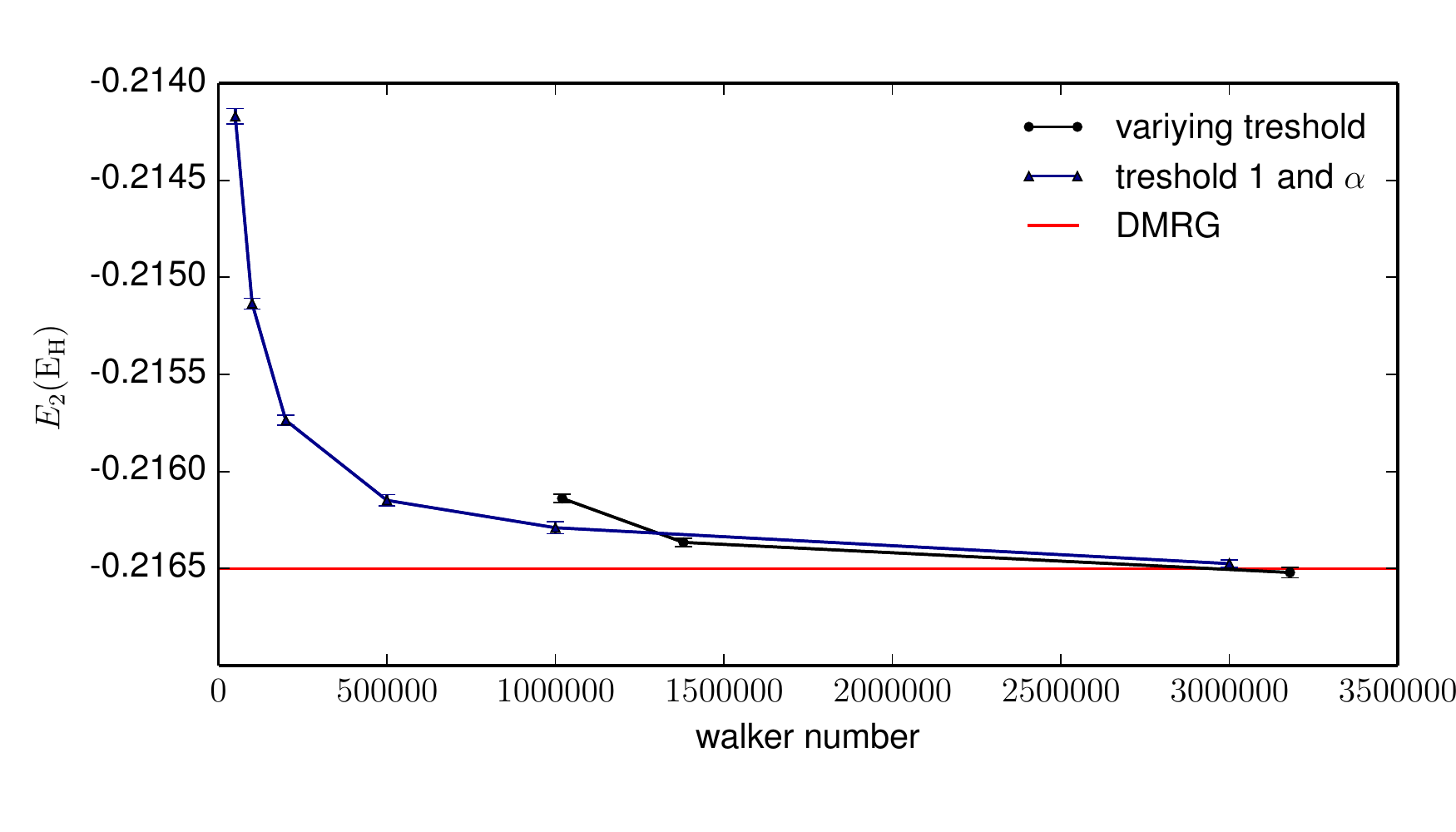}\caption{Comparison of the E2 energy with respect to the number of walkers.
The black curve is obtained by setting the initiator threshold to
different values, respectively 3,1 and no initiator threshold. The
blue curve is obtained by setting the initiator threshold to 1 and
controlling the number of walkers on $\protect\Ket{\Psi_{1}}$ by
using the $\alpha$ controlling parameter. For comparison purpose,
the value obtained using DMRG is in red.\label{fig1:compinit-alpha}}
\end{figure}

In Fig.\ref{fig1:compinit-alpha} we plotted the second order energy
for the Carbon dimer with the cc-pVQZ basis computed with the QMC-LCC
framework. The black line correspond to the simulations presented
in fig.\ref{fig1:swalknum} i.e. 50k walkers and a initiator threshold
of 3 in replica 0 and an initiator criterion of 1, 3 and no initiator
approximation for the first order response function. It can be seen
that without approximation, the computed energy is in agreement with
the one computed using MPS-LCC, shown in blue for reference. When
the initiator approximation is used, we obtain a second order energy
that is slightly higher than the correct one. However the estimation
remains quite good since the error in the energy is lower than 1 $mE_{H}$
with the two criteria used here. 

In Fig.\ref{fig1:compinit-alpha}, the blue curve is obtained with
an initiator threshold of 1 and different values of $\alpha$ to constrained
the number of walkers. The most interesting feature is that, for the
same number of walkers on $\Ket{\Psi_{1}}$ it seems to be better
to actually set a smaller initiator threshold and to use the $\alpha$
trick than increasing the initiator threshold. For instance for roughly
1M walkers the value obtained with initiator threshold of 1 and an
$\alpha$ value of $\approx0.73$ is 0.16 $mE_{H}$ lower than the
one obtained by using an initiator of 3.

In order to improve further the efficiency of our implementation,
we notice that the ${\cal H}_{1}$ space can be split as the orthogonal
sum of 8 smaller subspaces corresponding to the 8 subclasses of excitation
defined by Malrieu and collaborators\cite{angeli_introduction_2001,angeli_n-electron_2002}.
\begin{equation}
{\cal H}_{1}={\cal H}_{\left(-2,2,0\right)}\bigoplus{\cal H}_{\left(0,-2,2\right)}\bigoplus{\cal H}_{\left(-2,1,1\right)}\bigoplus{\cal H}_{\left(-2,0,2\right)}\bigoplus{\cal H}_{\left(-1,1,0\right)}\bigoplus{\cal H}_{\left(-1,0,1\right)}\bigoplus{\cal H}_{\left(-1,-1,2\right)}\bigoplus{\cal H}_{\left(0,-1,1\right)},\label{eq:H1=00003DSum_H}
\end{equation}

where the subscript described the change in number of electrons in
the core, active and virtual space with respect the ${\cal H}_{0}$
determinants. To each of this subspace correspond a subclass of excitation
that we can denote $\hat{V}_{(a,b,c)}$, connecting $\Ket{\Psi_{0}}$
to an ${\cal H}_{(a,b,c)}$ subspace.

When $\hat{H}_{0}$ is applied to a determinant belonging to one of
those subspaces, the generated determinants also belong to this subspace.
This means that during the dynamic, there are no interactions between
walkers belonging to 2 different subclasses. Instead of running a
single calculation by applying the full $\hat{V}$ to $\Ket{\Psi_{0}}$
it is possible to run 8 independent simpler calculations using each
of 8 different classes of excitations $\hat{V}_{(a,b,c)}$ and finally
sum up the 8 different second order energies obtained.

\section{Results\label{sec:Results}}

The LCC-QMC method proposed here has been applied to several organic
molecules. First we look at the behavior of the method with respect
to the size of the inactive space by computing the $E_{2}$ energy
for the $C_{2}$ molecule with different basis set. In each case the
active space consist of the valence orbitals of the molecule which
correspond to a (8,8) CAS. In addition to the CAS, there are 2 core
orbitals and 16, 50 and 100 virtual orbitals respectively in the cc-pVDZ,
cc-pVTZ and cc-pVQZ basis sets used here. The CASSCF orbitals have
been generated using the Molpro quantum chemistry package\cite{werner_molpro:_2012}.

We used 50k walkers and an initiator threshold of 3 in replica 0,
this threshold is set to 1 in replica 1. The response wavefunction
is sampled by applying the full $\hat{V}$ operator to the $\Ket{\Psi_{0}}$
wavefunction.

\begin{figure}
\centering{}\includegraphics[width=1\textwidth]{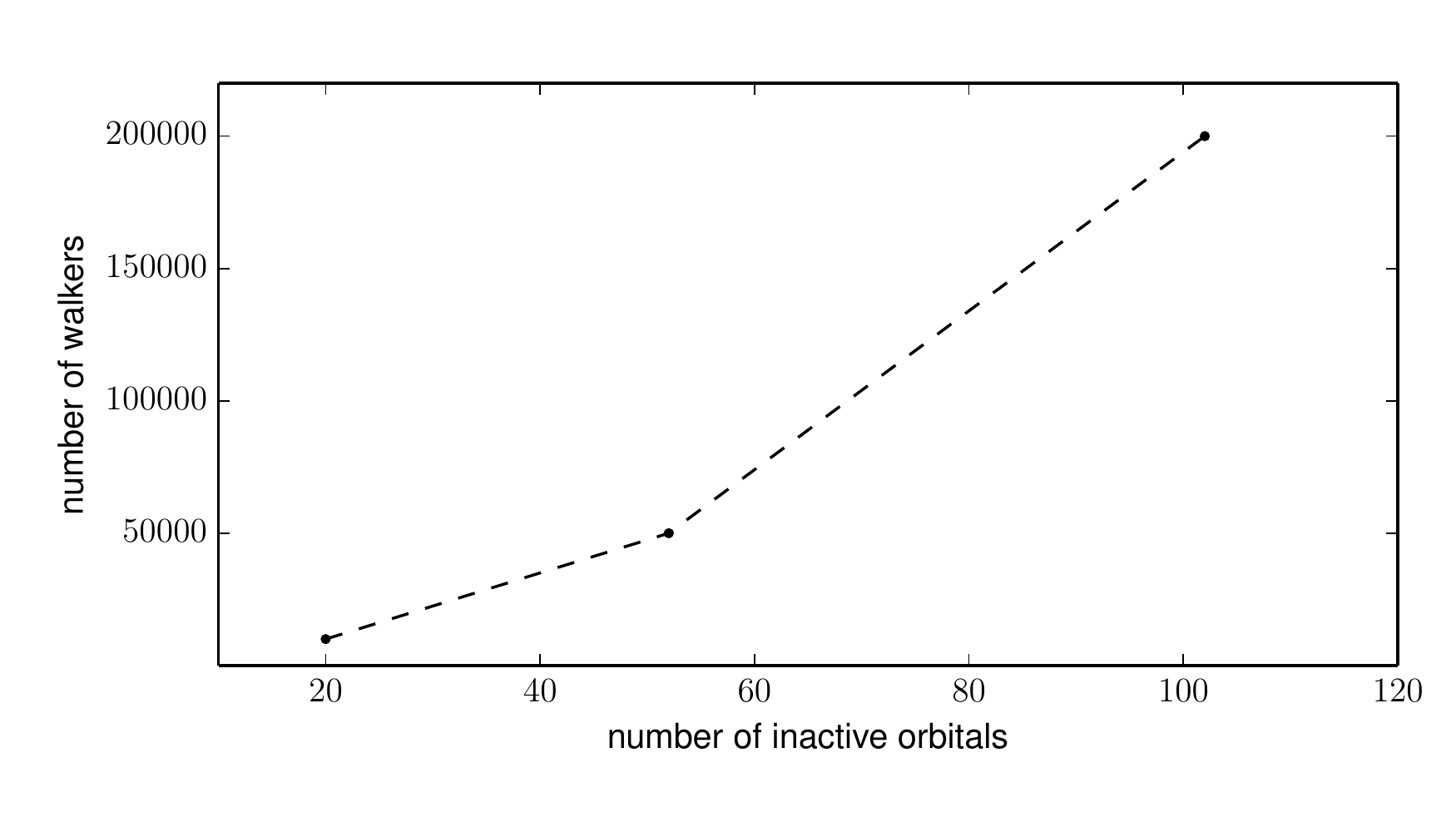}\caption{Comparison of the number of walkers that is necessary to use on replica
1 to reach a value within $1\ mE_{H}$ for $E_{2}$ with respect to
the energy computed by MPS-LCC used as reference. In all case the
CAS space is (8,8), the three different basis set used are cc-pVDZ
with 20 inactive orbitals, cc-pVTZ with 52 inactive orbitals and cc-pVQZ
with 102 inactive orbitals. The dotted line is here for eyes guidance.\label{fig1:enerc2basus}}
\end{figure}

To investigate how the cost of the response scale with the size of
the inactive space, we increase the number of walkers in replica 1
until the computed value of $E_{2}$ agrees within $1\ mE_{H}$ with
the same quantity computed deterministically with MPS-LCC using the
Block code\cite{sharma_spin-adapted_2012}. The Required number of
walkers are represented in Fig.\ref{fig1:enerc2basus}. From this
curve it can be seen that the number of walkers on $\Ket{\Psi_{1}}$
necessary to reach a $mE_{H}$ precision scales roughly as the square
of the number of inactive orbitals.

To test the applicability of the method, we now turn our attention
to the computation of benzene singlet-triplet gap. 

We used the same geometry than Roos et al\cite{matos_casscfcci_1987,roos_towards_1992},
i.e. C-C and C-H bond lengths of 1.395 $\textrm{Å}$ and 1.085 $\textrm{Å}$
and an hexagonal symmetry, for the singlet ground state and triplet
excited state . The ground state of the benzene molecule is singlet
$^{1}A_{1g}$, and we target the lower excited triplet of symmetry
$^{3}E_{1u}$. We used the Dunning cc-pVDZ basis set, the active space
contains 6 electrons and the six valence $\pi$ orbitals extended
with the six second shell $\pi$ orbitals. We used CASSCF orbitals
generated using Molpro.

For the $\Ket{\Psi_{0}}$ CASCI wavefunction we used an initiator
threshold of 3 and 100k walkers. In order to make the calculation
more tractable, we run 8 subcalculations corresponding to each of
the orthogonal classes of excitation $\hat{V}_{(a,b,c)}$. The initiator
threshold on $\Ket{\Psi_{1}}$ is set to 1.5. We start by using 100k
walkers for each first order function, but to test the applicability
of the methods, this number has been increased until the second order
energy agrees with a precision of 1 $mE_{h}$ with the one predicted
deterministicaly by LCC-MPS. The values obtained for the second order
energies for the different classes and the number of walkers it was
necessary to use on $\Ket{\Psi_{1}}$ to obtain these values are specified
in Table.\ref{tab:-energies-benz}. The $\left(-1,1,0\right)$ class
is not contributing since it has no overlap with the $\Ket{\Psi_{0}}$
wavefunction for spatial orbital symmetry reasons.

We can see by looking at this table that the classes of excitations
are not equivalent in terms of the number of walkers necessary to
converge. The classes involving virtual orbitals require more walkers
to reduce the initiator error. In particular the $\left(-2,0,2\right)$
class is particularly difficult, this is understandable since in this
case this the class containing the biggest number of determinants.
Considering the Singlet-Triplet gap, the CASSCF value is equals to
4.97 eV while it is reduced to 4.88 eV when the LCC correction is
used, the experimental value determined by electron-impact spectroscopy
is 4.76 eV\cite{doering_lowenergy_1969}. We also computed the singlet-triplet
gap with CASPT2 where internal contraction is used only for the subspaces
requiring at most the knowledge of the second order reduced density
matrix while the other subspaces are left uncontracted and NEVPT2
strongly contracted using Molpro, we obtained an difference of 4.65
eV and 5.03 eV respectively. 

\begin{table}
\begin{centering}
\begin{tabular}{|c|c|c|c|c|c|}
\hline 
 & \multicolumn{1}{c}{} & Singlet &  & \multicolumn{1}{c}{} & Triplet\tabularnewline
\cline{1-3} \cline{5-6} 
Class & $E_{2}$ & Walkers number &  & $E_{2}$ & Walkers number\tabularnewline
\hline 
\hline 
$\left(-2,2,0\right)$ & -0.02025  & 500 000 &  & -0.01167  & 500 000\tabularnewline
\cline{1-3} \cline{5-6} 
$\left(0,-2,2\right)$ & -0.01716 & 500 000 &  & -0.01681 & 500 000\tabularnewline
\cline{1-3} \cline{5-6} 
$\left(-2,1,1\right)$ & -0.03330 & 1000 000 &  & -0.03558 & 1000 000\tabularnewline
\cline{1-3} \cline{5-6} 
$\left(-2,0,2\right)$ & -0.46096  & 10 000 000 &  & -0.46706 & 10 000 000\tabularnewline
\cline{1-3} \cline{5-6} 
$\left(-1,1,0\right)$ &  0.00000 & N/A &  &  0.00000 & N/A\tabularnewline
\cline{1-3} \cline{5-6} 
$\left(-1,0,1\right)$ &  -0.21671 & 1000 000 &  & -0.20215 & 1000 000\tabularnewline
\cline{1-3} \cline{5-6} 
$\left(-1,-1,2\right)$ & -0.08402 & 1000 000 &  &  -0.09871 & 5000 000\tabularnewline
\cline{1-3} \cline{5-6} 
$\left(0,-1,1\right)$ &  -0.00669 & 500 000 &  &  -0.00680 & 500 000\tabularnewline
\hline 
CASCI & -230.8070 &  &  & -230.6244 & \tabularnewline
\cline{1-3} \cline{5-6} 
CASCI+LCC & -231.6461 &  &  & -231.4667 & \tabularnewline
\cline{1-3} \cline{5-6} 
\end{tabular}\caption{$E_{2}$ energies in $E_{H}$ for the 8 different classes of excitation
for the singlet (left) and triplet (right) predicted by MPS-LCC and
LCC-QMC. Next to each value is the number of walkers necessary to
get this number with the method described in this article. The $\left(-1,1,0\right)$
class has no overlap with $\protect\Ket{\Psi_{0}}$ because the excitation
are forbidden by symmetry. \label{tab:-energies-benz}}
\par\end{centering}
\end{table}

We then turned our attention to the computation of the triplet-singlet
gap in the m-xylylene diradical. The study of radical organic species
and the prediction of their spin properties is of interest because
of potential application in the developement of molecule-based magnetic
materials\cite{miller_organic_1994}.

The key parameter for such application is the triplet-singet gap,
thus some effort have been done in order to tune this parameter. Along
the experiments it is interesting to predict the value of this singlet
triplet gap to guide the synthesis of new promising molecules.

Among the different organic diradicals, the m-xylylene has been used
quite often as a benchmark system since it is rather stable and quite
well characterized experimentally. The molecule belong to $C_{2v}$
point group, the ground state has been proved to be a triplet by using
EPR\cite{wright_electron_1983}, and it has electronic state $^{3}B_{2}$.
It has been shown by using NIPES that the lowest lying excited state
is $^{1}A_{1}$\cite{wenthold_photoelectron_1997}. This system has
been quite extensively studied numerically. For instance Mañeru \textit{et
al}\cite{reta_maneru_tripletsinglet_2014}\textit{ }carried an extensive
study with DFT and several wavefunction methods exploring the effect
of the geometries and of the basis set on the predicted gap. They
found that the choice of the basis set have low influence on the predicted
gap, however as expected the value of the gap is highly dependent
of the choice of the functional. On the other hand, all the different
wavefunctions methods they used tends to overestimate the singlet
triplet gap. Even with more sophisticated basis set their predicted
value of the singlet-triplet gap remains quite close to the value
of 4092 $cm^{-1}$ previously predicted by Hrovat et al using a CAS-PT2
calculation with a CAS (8,8) and 6-31g{*} basis set\cite{hrovat_effects_1998}.

As the authors states ``the difficulty of the wave function-based
methods in describing the triplet\textminus singlet gap arises quite
unequivocally from dynamical correlation''. They proposed to extend
the CAS space as a way to recover the dynamical correlation is improperly
taken account, however as they correctly stated this will make the
calculation computationally challenging.

As an alternative we decided to examine how a better treatment of
dynamical correlation by using the LCC multireference perturbation
theory with the same CAS space would improve the prediction of the
singlet-triplet gap. We used the equilibrium geometries optimized
at the CASSCF(8,8)/6-311++g{*}{*} level given in supporting information
of Mañeru et al, for both the singlet and the triplet. We used the
same basis set to run a CASSCF(8,8) calculation to generate the 2
and 4 indexes integrals files with the Molpro quantum package. We
computed the $E_{2}$ energies for the 8 classes with MPS-LCC using
internal contraction as a reference. We start by using a procedure
similar to the one used with the benzene molecule which is setting
the iniator criteria to 1.5 in for the first order response functions
and progressively increasing the number of walker sample it. However
except for the two classes $\left(-2,2,0\right)$ and $\left(0,-1,1\right)$
this procedure has failed, since the obtained value where much higher
than the one predicted by MPS-LCC. We thus decided to run calculation
with a initiator criterium of 1 and no control of the population i.e
an $\alpha$ factor of 1. The number obtained and the number of walkers
reached on replica 1 are given in table \ref{tab:-energies-xyl}.
With this procedure the obtained number are in agreement with the
MPS-LCC results, there are usually a bit more negative since this
is a fully uncontracted technique. However this approach is not really
practical since the number of walkers reached on $\Ket{\Psi_{1}}$
is generally huge, making the calculation extremely costly. Moreover
in the case of the $\left(-2,0,2\right)$ and $\left(-1,-1,2\right)$
classes it was not possible to do the calculation. 

\begin{table}
\centering{}%
\begin{tabular}{|c|c|c|c|c|c|}
\hline 
 & \multicolumn{1}{c}{} & Singlet &  & \multicolumn{1}{c}{} & Triplet\tabularnewline
\cline{1-3} \cline{5-6} 
Class & $E_{2}$ & Walkers number &  & $E_{2}$ & Walkers number\tabularnewline
\hline 
\hline 
$\left(-2,2,0\right)$ & -0.00625  & 500 000 &  & -0.00630 & 500 000\tabularnewline
\cline{1-3} \cline{5-6} 
$\left(0,-2,2\right)$ & -0.0324 & 36 000 000 &  & -0.0319 & 650 000 000\tabularnewline
\cline{1-3} \cline{5-6} 
$\left(-2,1,1\right)$ & -0.0438 & 50 000 000 &  & -0.0437 & 40 000 000\tabularnewline
\cline{1-3} \cline{5-6} 
$\left(-2,0,2\right)$ & -0.46096 {*} &  &  & -0.46706{*} & \tabularnewline
\cline{1-3} \cline{5-6} 
$\left(-1,1,0\right)$ &  0.00000 & N/A &  &  0.00000 & N/A\tabularnewline
\cline{1-3} \cline{5-6} 
$\left(-1,0,1\right)$ &  -0.182 & 13 000 000 &  & -0.180 & 88 000 000\tabularnewline
\cline{1-3} \cline{5-6} 
$\left(-1,-1,2\right)$ & -0.08402{*} &  &  &  -0.09871{*} & \tabularnewline
\cline{1-3} \cline{5-6} 
$\left(0,-1,1\right)$ &  -0.0168 & 500 000 &  &  -0.0152 & 500 000\tabularnewline
\hline 
CASCI & -230.8070 &  &  & -230.6244 & \tabularnewline
\cline{1-3} \cline{5-6} 
CASCI+LCC & -231.6461 &  &  & -231.4667 & \tabularnewline
\cline{1-3} \cline{5-6} 
\end{tabular}\caption{$E_{2}$ energies in $E_{H}$ for the 8 different classes of excitations
for the singlet (left) and triplet (right) of the m-xylylene diradical
predicted by LCC-QMC. Next to each value is the number of walkers
necessary to get this number with the method described in this article.
It has not been possible to make the classes $\left(-2,0,2\right)$
and $\left(-1,-1,2\right)$ converge with LCC-QMC, the $E_{2}$ labeled
({*}) are obtained with MPS-LCC using internal contraction \label{tab:-energies-xyl}}
\end{table}

If we retains the values predicted by MPS-LCC for the classes where
QMC-LCC cannot be converged, we found a value of 3440 $cm^{-1}$ for
the triplet-singlet gap of m-xylylene, a value that agrees well with
the experimental value $3358\pm70\ cm^{-1}$. This shows that using
the MRLCC perturbation theory improves a lot the description of the
dynamical correlation for this system.

Thus as Mañeru\textit{ et al} stated, the problem in the description
of the m-xylylene was a proper treatment of the dynamical correlation.
However this problem is still out of reach for the fully uncontracted
LCC-QMC, this indicated clearly that to make LCC-QMC practical it
would be necessary to implement some internal contraction treatment
of the perturbation classes. However he fact that the most easy classes
in the LCC-QMC approach is the one that would require the 3 and 4-RDM
to be computed with internal contraction is encouraging.

\section{Conclusions\label{sec:Conclusions}}

In this article we described a way to do CASCI+MRLCC calculation that
is completely stochastic. The zero order wavefunction is computed
with the FCIQMC approach, \textit{i.e} a population of signed walkers
which is evolving thanks to a series of stochastic rules, is sampling
the CASCI wavefuntion and solving the zero order Hamiltonian eigenproblem.
Simultaneously, a second population of walkers submitted to a different
set of stochastic process is sampling the first order wavefunction,
by finding the steady state of a differential equation. The first
order wavefunction being a function of the zero order one, we include
a source term in the population dynamic sampling $\Ket{\Psi_{1}}$
that depend of the population in $\Ket{\Psi_{0}}$, this requires
spawning from one replica to the other, and this is the main originality
of the proposed algorithm. We presented different strategy to make
the use of this technique practicable, some are directly adapted from
the strategies proposed for FCIQMC, such as the initiator approximation
and the semi-stochastic approach. We also proposed to scale the source
term as a way to control the population in the replica sampling the
first order wavefunction. Because the size of the CASCI and of the
perturbation are quite different it was also necessary to have different
timesteps for the 0 to 0, 0 to 1 and 1 to 1 spawning steps. We illustrate
the possibility of the proposed methods on several applications. First
we shows that this approach allows to recover the results computed
by the deterministic MPS-LCC method on the case of the $C_{2}$ molecule.
This calculation shows that the cost of the calculation, linked to
the number of walkers, scale quadratically with the number of inactive
orbitals.

The computation of the singlet-triplet gap of the challenging m-xylylene
diradical with the proposed method confirmed the better performance
of the MRLCC perturbation theory with respect to the CASPT2 technique.
It has been possible to obtain a value of this gap that is in good
agreement with the experimental results while the CASPT2 approach
using the same active space overestimate this quantity by 20\%. However
it has also been demonstrated that the proposed algorithm is still
not adapted to study such complicated systems since the class of excitation
involving two cores electrons and to virtuals holes cannot be computed
for the m-xylylene molecule. This problem has been circumvent by computing
those classes by using the contracted MPS-LCC technique. This point
out the necessity to develop a similar contracted approach with our
fully stochastic procedure, this is currently under investigation.

The LCC-QMC approach proposed here would allow to treat both large
active space and large inactive space; this associated with the recently
demonstrated possibility of using FCIQMC-CASSCF\cite{li_manni_combining_2016}
will allow to study systems that are currently out of reach.

\section*{Acknowledgments}

The calculations made use of the facilities of the Max Planck Society\textquoteright s
Rechenzentrum Garching. We are grateful to Pr. Illas and Dr. Daniel
Reta for providing us their equilibrium geometries for the m-xylylene
molecule.

\selectlanguage{french}%
\bibliographystyle{apsrev}
\bibliography{biblioLCC}
\selectlanguage{english}%

\end{document}